\documentclass[aps,12pt,prl]{revtex4-2}

\usepackage{bm}

\usepackage{natbib}

\usepackage{hyperref} 
\hypersetup{
    colorlinks = true,
    linkcolor={gray},
    citecolor={blue!50!black},
    urlcolor={blue!50!black}
}
\usepackage{siunitx}
\usepackage{mathtools}
\usepackage{bbold}
\usepackage[normalem]{ulem}
\usepackage{enumerate}
\usepackage[dvipsnames]{xcolor}
\def\be{\begin{equation}}
\def\ee{\end{equation}}
\def\e#1{\label{#1}\end{equation}}
\def\bea{\begin{eqnarray}}
\def\eea{\end{eqnarray}}
\def\ea#1{\label{#1}\end{eqnarray}}

\def\bem#1{\begin{mathletters}\label{#1}}
\def\eml{\end{mathletters}}

\newcommand{\ket}[1]{\left| #1 \right>}

\def\4#1{{\boldsymbol{#1}}}
\def\8#1{{\widetilde{#1}}}
\def\bse{\begin{subequations}}
\def\ese{\end{subequations}}

\begin{document}

\title{Matters Arising: Distributed quantum sensing with mode-entangled spin-squeezed atomic states}

\author{Liam P. McGuinness}
\affiliation{Laser Physics Centre, Research School of Physics, Australian National University, Acton, Australian Capital Territory 2601, Australia \\ \textnormal{Email: \href{mailto:liam@grtoet.com}{liam@grtoet.com}} }

\begin{abstract}
In ``Distributed quantum sensing with mode-entangled spin-squeezed atomic states" Nature (2022) \cite{Malia2022}, Malia et.~al. claim to improve the precision of a network of clocks by using entanglement. In particular, by entangling a clock network with up to four nodes, a precision 11.6\,dB better than the quantum projection noise limit (i.e. precision without any entanglement) is reported. These claims are incorrect, Malia et.~al. do not achieve an improved precision with entanglement. Here we show their demonstration is more than two orders of magnitude worse than the quantum projection noise limit.
\end{abstract}

\maketitle

\pagestyle{plain}

The central message in ``Distributed quantum sensing with mode-entangled spin-squeezed atomic states" Nature (2022) \cite{Malia2022} is that by entangling atoms in an atomic clock network, a precision is demonstrated that is impossible to attain using the same number of atoms and time without entanglement. Should we accept this message? Putting aside the impressive technical achievements in the paper, we can objectively assess whether the experimental data are in agreement with the claim.

The final two figures of the paper present the supporting data. In Fig.\,3, $\Delta (\bar{\theta})$ is plotted as a function of the number of clocks $M$, each with $N = 45,000$ atoms, where the black line ($1/\sqrt{M N}$) is supposed to denote a limit that cannot be surpassed without entanglement -- the quantum projection noise limit (QPN). One point should be made clear, $\Delta (\bar{\theta}) > 1/\sqrt{M N}$ is very definitively not the limit without entanglement. To see this requires an explanation of what $\Delta (\bar{\theta})$ is. Despite the authors calling $\Delta (\bar{\theta})$ the `measured sensitivity', it is not a sensitivity at all. $\bar{\theta}$ is simply the difference in values between two measurements (for $M = 1$). Taking the square-root of the variance of this difference we obtain $\Delta (\bar{\theta})$. A low value of $\Delta (\bar{\theta})$ implies that the difference between the two measurements remains the same upon repetition. Now there are many ways that the difference between two measurements can be stable without requiring entanglement. One very obvious method is to measure spin population along $z$ and then measure the $z$ population again. For precise measurements limited by quantum projection noise, a value of $\Delta (\bar{\theta})$ arbitrarily close to zero can be obtained. There is no need for entanglement here, just the ability to perform a quantum non-demolition (QND) measurement. Furthermore, $\Delta (\bar{\theta})$ is not even bound by quantum projection noise, a low variance also results from poor measurements producing no data, since the variance of a sequence of nulls is zeros, and we do not even need to perform a QND. Importantly, if one just wants to minimise $\Delta (\bar{\theta})$ there is no requirement that $\bar{\theta}$ be proportional to something we want to measure, nor that the variance reflect our uncertainty in this quantity.

Somewhat less obvious is the sequence of measurements Malia et. al. perform. After initialising the atoms in a given spin state (e.g. spin up $\ket{\uparrow}$), a $\pi/2$-rotation creates an equal superposition of $\ket{\uparrow}$ and $\ket{\downarrow}$, before a QND measurement projects the atoms. A microwave spin-echo $\pi/2 - \pi - \pi/2$ is then performed before a second QND measurement readouts the population. Using beautifully ambiguous terminology the authors mention that the phase of the microwave pulses are adjusted `so that the $J_z$ distributions are in metrologically sensitivity configurations', i.e. adjusted to obtain a spin population close to the first measurement. Again there is no need for entanglement here and the projection noise limit for the variance is zero.

The claims in the abstract of `4.5 decibels better precision than one without spatially distributed entanglement, and 11.6 decibels improvement as compared to a network of sensors operating at the quantum projection noise limit' come directly from this dataset. We can state unequivovally, this is incorrect on several accounts. Firstly, no analysis of the measurement precision is presented -- just noise. Secondly the limit surpassed by the authors can be surpassed by a network of unentangled sensors operating at the quantum projection noise limit (or even one operating far from this limit). Finally, the authors do not present any evidence that the atoms are entangled. Malia et. al. claim to generate entanglement using a projective QND measurement. Evidence of entanglement is presented in the form of measurement statistics below the quantum projection noise limit. We note that projective measurement is a common technique to initialise atoms in an eigenstate, so this technique does not definitively produce entanglement. And as we have shown, the observations presented by the authors are fully consistent with separable states.

In Fig.\,4, the authors present the measured phase uncertainty obtained with 220,000 entangled (green) and non-entangled (black) atoms. This uncertainty is inferred from differences between two measurements, $\bar{\theta}$ as described above, where `entanglement' is generated with a projective measurement. To reiterate, $\Delta \bar{\theta}$ plotted in Fig.\,4b) is not the measurement precision, it is just an analysis of the measurement noise. The authors appear to be aware of this since they shift between terminology in the paper. Take this sentence in the introduction: ``This entanglement enhances the \textbf{precision} of the frequency comparison within networks of identical clocks, each containing 45,000 atoms per mode." Compare with the next sentence: ``A mode-entangled four-mode network exhibits \textbf{noise} roughly 4.5(0.8) dB lower than that of an equivalent mode-separable network of spin-squeezed states (SSS) and 11.6(1.1) dB lower than a network of coherent spin states (CSS) operating at the QPN limit..." (emphasis added). Of course there are many strategies to reduce the noise below the values demonstrated by Malia et. al. without using entanglement. The interesting question is which of these strategies lead to improved precision.


To answer to this question, where the phase $\phi$ of the Raman laser is inferred from measurements of $\bar{\theta}$, requires two important pieces of additional information. \textbf{1)} $\partial \bar{\theta}/\partial \phi$ at the value of $\bar{\theta}$ where the noise is characterised. I.e. the response to the parameter of interest, which \emph{together} with the noise defines the statistical uncertainty. \textbf{2)} Accurate knowledge of $\phi$ as a function of $\bar{\theta}$ (at least at $\bar{\theta} \approx 0$), ideally ensuring the correspondence is one-to-one. I.e. the systematic uncertainty in $\phi$. The authors do not present any of this information, and despite requests, they have not yet provided any data for analysis. In Fig.\,4c) a dataset is provided that can be used to estimate the phase uncertainty for unentangled atoms. From the fit to the data an uncertainty $\Delta \phi = 0.08$\,rad is obtained (per data-point and excluding the first data-point as an outlier). This is more than an order of magnitude worse than than any uncertainty shown in Fig.\,4b), highlighting that $\Delta \bar{\theta}$ is not the uncertainty in estimating the laser phase. Given the experimental parameters for the mode-entangled experiment, we estimate that the uncertainty in estimating the laser phase for this dataset is more than two-orders of magnitude worse than quoted.

One might ask, how do Malia et. al. observe a lower value of $\Delta (\bar{\theta})$ with `entangled' atoms compared to unentangled atoms? The answer is simple. With unentangled atoms Malia et. al. present $\Delta (\bar{\theta})$ near the point of maximum QPN. This also happens to be the most sensitive point of operation for a quantum interferometer, since the signal response is simultaneously maximised \cite{Itano1993}. With entangled atoms, Malia et. al. present $\Delta (\bar{\theta})$ for values of $\bar{\theta}$ very close to zero, where it is minimised. The lesson here is that lower noise does not necessarily correspond to better sensivity, and for devices limited by QPN, the best sensitivity may occur at the point where noise is greatest.

A final point, an uncertainty $\Delta \phi > 1/\sqrt{N}$ does not limit the precision $N$ unentangled atoms can estimate the phase of a near-resonant field, e.g. Raman laser. First of all, no law prevents us performing more than one measurement per atom (note Malia et. al. perform two measurements) to reduce the uncertainty below this level. Secondly, even restricting analysis to one measurement per atom, the above uncertainty limit only holds assuming a one-to-one correspondence between the atomic phase and the phase of the Raman laser \cite{Wootters1981}. By increasing the interaction time it is possible for a small shift in phase of the Raman laser to generate an arbitrarily large atomic phase, thus surpassing this limit. The following bound does however limit the uncertainty in phase estimation with unentangled spins \cite{McGuinness2021}:
\be
\Delta \phi > \frac{1}{\sqrt{N}\Omega_R T},
\ee
where $\Omega_R$ is the Rabi frequency of interaction with the Raman laser given we have no prior knowledge on $\phi$. The lesson here. Only by taking into account how long the measurement takes, can one obtain a strict precision limit.

For the data in Fig.\,4 the Raman $\pi$-pulse took 2\,$\mu$s and we have $\Omega_R = 2\pi\times 250$\,kHz. With a total experimental duration $>$\,7\,ms, the projection noise limit is $\Delta \phi > 50\,\mu$rad, much lower than the limit that Malia et. al. compare to. Excluding the 7\,ms interval and only accounting for the finite duration of the microwave and laser pulses, the projection noise limit is 0.2\,mrad, again more than an order of magnitude lower than the value Malia et. al. compare to, where we have not included the time to readout or prepare the spins. Using the correct QPN limit as a benchmark, Malia et. al. demonstrate a precision more than three orders of magnitude worse using `entangled' spins. In other words, a single atom would suffice to outperform the demonstrated precision.

Malia et. al. use a variety of techniques to give the impression of surpassing the unentangled precision limit without actually doing so. This includes analysing noise instead of precision, using an incorrect definition of the quantum projection noise limit and not including the measurement time. These strategies are common in the field, but they are not the only examples of incorrect comparisons. One point though does holds universally across the entire field of quantum metrology -- when correctly defined, no experiment has ever surpassed the precision limit of unentangled particles.

\section{Communication with Authors and Nature editor}
\noindent A request for the data used in plotting Fig.\,4b) was sent to the authors on Jan 10, 2023. As of writing, the authors have not provided this data. A copy of this critique was sent to the authors on Jan 27, 2023. A response was received but there was no comment on the correctness of the critique.\\

\noindent On Feb 1, 2023, this critique, supplemented by correspondence with the authors, was submitted to \emph{Nature} to be published as a Matters Arising. On Feb 9, 2023 the manuscript was rejected.


\begin{thebibliography}{4}
\expandafter\ifx\csname url\endcsname\relax
  \def\url#1{\texttt{#1}}\fi
\expandafter\ifx\csname urlprefix\endcsname\relax\def\urlprefix{URL }\fi
\providecommand{\bibinfo}[2]{#2}
\providecommand{\eprint}[2][]{\url{#2}}

\bibitem{Malia2022}
\bibinfo{author}{Malia, B.~K.}, \bibinfo{author}{Wu, Y.},
  \bibinfo{author}{Martínez-Rincón, J.} \&
  \bibinfo{author}{Kasevich, M.~A.}
\newblock \bibinfo{title}{Distributed quantum sensing with mode-entangled spin-squeezed atomic states}.
\newblock \emph{\bibinfo{journal}{Nature}} \textbf{\bibinfo{volume}{612}},
  \bibinfo{pages}{661--665} (\bibinfo{year}{2022}).
\newblock \urlprefix\url{https://doi.org/10.1038/s41586-022-05363-z}.

\bibitem{Itano1993}
\bibinfo{author}{Itano, W.~M.} \emph{et~al.}
\newblock \bibinfo{title}{Quantum projection noise: Population fluctuations in two-level systems}.
\newblock \emph{\bibinfo{journal}{Physical Review A}}
  (\bibinfo{year}{1993}).
\newblock \urlprefix\url{http://link.aps.org/doi/10.1103/PhysRevA.47.3554}.

\bibitem{McGuinness2021}
\bibinfo{author}{McGuinness, L.~P.}
\newblock \bibinfo{title}{The case against entanglement improved measurement precision}.
\newblock \emph{\bibinfo{journal}{arXiv preprint}}
  (\bibinfo{year}{2021}).
\newblock \urlprefix\url{https://doi.org/10.48550/arXiv.2112.04354}.

\bibitem{Wootters1981}
\bibinfo{author}{Wootters, W.~K.}
\newblock \bibinfo{title}{Statistical distance and Hilbert space}.
\newblock \emph{\bibinfo{journal}{Physical Review D}}
  (\bibinfo{year}{1981}).
\newblock \urlprefix\url{https://link.aps.org/doi/10.1103/PhysRevD.23.357}.




%
%
%
%
%
%
%
%
%
%
%
%
%
%
%
%
%

\end{thebibliography}
\end{document}